\documentclass[pra,twocolumn,showpacs,preprintnumbers,amsmath,amssymb,superscriptaddress,aps,citeautoscript]{revtex4-1}

\usepackage{graphicx}
\usepackage{dcolumn}
\usepackage{bm}
\usepackage{epsfig}
\usepackage{amsmath}
\usepackage{amssymb}
\usepackage{color}
\usepackage{bbm}
\usepackage{braket} 

\newcommand{\im}{{\operatorname{i}}}
\newcommand{\ex}{{\operatorname{e}}}
\newcommand{\id}{{\operatorname{d}}}

\newcommand{\mycite}[1]{{\textsuperscript{\cite{#1}}}}

\usepackage[colorlinks=true,citecolor=blue,linkcolor=blue,urlcolor=blue]{hyperref}
\begin{document}

\title{Quantum Simulation of Non-perturbative Cavity QED with Trapped Ions}

\author{Tuomas Jaako}
\email{tuomas.jaako@tuwien.ac.at}
\affiliation{Vienna Center for Quantum Science and Technology, Atominstitut, TU Wien, 1040 Vienna, Austria}

\author{Juan Jos\'e Garcia-Ripoll}
\affiliation{Instituto de F\' isica Fundamental, IFF-CSIC, Calle Serrano 113b, Madrid E-28006, Spain}

\author{Peter Rabl}
\affiliation{Vienna Center for Quantum Science and Technology, Atominstitut, TU Wien, 1040 Vienna, Austria}

\date{\today}
\begin{abstract}
We discuss the simulation of non-perturbative cavity-QED effects using systems of trapped ions. Specifically, we address the implementation of extended Dicke models with both collective dipole-field and direct dipole-dipole interactions, which represent a minimal set of models for describing light-matter interactions in the ultrastrong and deep-strong coupling regime. We show that this approach can be used in state-of-the-art trapped ion setups  to investigate excitation spectra or the transition between sub- and superradiant ground states, which  are currently not accessible in any other physical system. Our analysis also reveals the intrinsic difficulty of accessing this non-perturbative regime with larger numbers of dipoles, which makes the simulation of many-dipole cavity QED a particularly challenging test case for future quantum simulation platforms.
\end{abstract}

\maketitle
%
%
\section{Introduction}
Quantum electrodynamics (QED) is our fundamental theory for describing the dynamics of charges coupled to the quantized electromagnetic field. In contrast  to quantum chromodynamics for the strong force, QED is a weakly-interacting theory, which is characterized by the small value of the electromagnetic finestructure constant, $\alpha_{\rm fs}\simeq 1/137 $. Apart from its implications in particle physics, this property  affects as well many processes relevant in our daily life, for example, the way in which light interacts with atoms, molecules and solid matter. Specifically, the smallness of $\alpha_{\rm fs}$ implies that the coupling strength $g$ between a single elementary dipole and a single photon of frequency $\omega_c$ is constrained to $g/\omega_c\lesssim \sqrt{2\pi\alpha_{\rm fs}}\ll1 $.\mycite{HarocheCQED,Devoret2007,debernardis2018} As a consequence, 
%
the coupling between matter and photons can typically be treated as a small perturbation on top of the absolute energy scales and does not considerably alter the overall structure of ground- and excited states.

In recent years there has been a growing interest in the physics of light-matter interactions beyond this conventional coupling regime.\mycite{forndiaz19,kockum19} In many experiments with dense ensembles of electrons, excitons or molecules it is now possible to reach ultrastrong-coupling (USC) conditions,\mycite{Ciuti2005} where the collective coupling, $G= \sqrt{N}g $, between $N\gg1$ dipoles and a single cavity mode reaches a considerable fraction of the bare photon frequency. Moreover, in the field of circuit QED,\mycite{Wallraff2004,Blais2004,Gu2017} analogue models for light-matter interactions can be implemented by coupling artificial atoms, i.e., superconducting two-level systems, with microwave photons. In this case the bound mentioned above can be overcome by using high-impedance resonators or galvanic coupling schemes, such that $g/\omega_c\gtrsim 1$ can be realized even with a single qubit.\mycite{forndiaz17,yoshihara17} In this, often called deep-strong-coupling (DSC), regime,\mycite{Casanova2010} the interaction between atoms and photons has a non-perturbative effect on the energy-level structure of the combined system.


Despite a considerable amount of work on this subject, non-perturbative effects in cavity and circuit QED are still little understood. In the past, a lot of theoretical studies in this field have been devoted to the quantum Rabi model and variations thereof.\mycite{forndiaz19} This model, however, only describes the coupling of a single dipole to a cavity mode and does not capture cavity-mediated interaction effects. In turn, collective interactions and phase transitions are usually discussed in the opposite limit of a large number of dipoles, $N\gg1$. In this case the relevant coupling parameter per atom, $g/\omega_c$, is typically assumed to be small, such that most effects can be understood in terms of conventional electrodynamics.\mycite{debernardis2018} Thus, the most intriguing regime, where both non-perturbative and many-body effects play a role, remains largely unexplored, which is also related to the fact that the combined conditions $g/\omega_c \gtrsim 1$ and $N>1$ have not been demonstrated in any of the mentioned experimental platforms so far. This motivates the search for alternative quantum simulation schemes, where the USC physics can be explored, independently of any technological constraints and under fully controlled conditions.\mycite{Dimer2007,Ballester2012,Zou2014,Pedernales2015,Puebla2016,Schneeweiss2018,aedo2018}

In this work we investigate the use of systems of trapped ions as a quantum simulator for multi-dipole  cavity QED systems in the USC regime. This platform is naturally suited for this purpose since experimental techniques for implementing Jaynes-Cummings-,  Rabi- and Dicke-type couplings between the internal atomic states and motional modes (which represent the photons in the effective model) are already well-established.\mycite{Leibfried2003,Lv2017,SafaviNaini2018,Cohn2018} However, for $N>1$ such models provide very restricted or inconsistent descriptions of quantum electrodynamics beyond the weak-coupling regime and must be complemented by additional interaction terms.\mycite{debernardis2018} This includes all-to-all and short-range dipole-dipole interactions, which account for the  ``$P^2$-term" \mycite{PhotonsAndAtoms,Todorov2012,Todorov2014} and electrostatic forces, respectively. Both contributions play a dominant role for cavity QED systems in the regime of very large coupling. Here we describe how such extended Dicke models can be simulated in a chain of trapped ions with a large tunability of the effective model parameters. Specifically, we demonstrate that characteristic USC effects in the excitation spectra and in the ground state of this system can be probed with $N\approx 2-10$ ions in state-of-the-art experimental setups. Therefore, small-scale trapped-ion quantum simulators can already be used to explore the few-dipole USC regime of cavity QED, which is not accessible in any other physical platform today. 


\section{Ultrastrong-coupling cavity QED}

\begin{figure}[t]
    \begin{center}
        \includegraphics[width=\columnwidth]{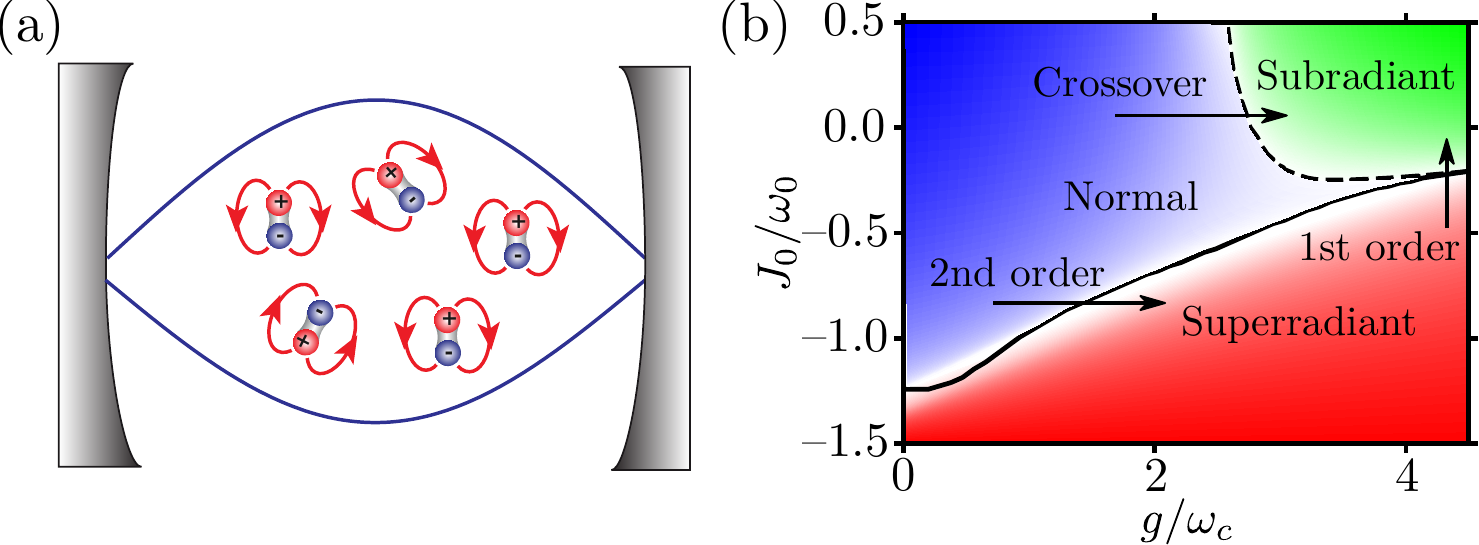}
        \caption{(a) Illustration of a prototypical cavity QED setup, where multiple two-level dipoles are coupled to a single dynamical electromagnetic mode and among each other via direct dipole-dipole interactions. (b) Outline of the ground state phase diagram of the extended Dicke model defined in Eq.~\eqref{eq:h_edm} under the collective spin approximation, $J_{ij}=J_0/N$ and $N=8$. Note that compared to Ref.\ \cite{debernardis2018}, here an independent scaling of $J_0$ and $g$ is assumed.}
        \label{Fig1:CavityQED}
    \end{center}
\end{figure}

Figure \ref{Fig1:CavityQED}(a) depicts a generic cavity QED setup, where $N$ two-level dipoles with transition frequency $\omega_0$ are coupled to a single electromagnetic mode of frequency $\omega_c$. Under the assumption that the electric field of the dynamical mode is sufficiently homogeneous, such a scenario is described by the Hamiltonian ($ \hbar = 1 $)\mycite{debernardis2018}
\begin{equation}\label{eq:h_edm}
\begin{split}
    H_{\rm cQED} = &\omega_{c}a^{\dagger}a + \omega_0 S_z+g (a^{\dagger} + a)S_x\\
    &+\dfrac{ g^2}{ \omega_{c}} S_x^2  +  \frac{1}{4}\sum_{i,j=1}^N   J_{ij} \sigma_i^x\sigma_j^x.    
\end{split}
\end{equation}
Here the $\sigma^\alpha_i$, where $ \alpha = x,y,z $, are the Pauli-operators for the $i$-th dipole, $S_\alpha=1/2\sum_i\sigma^\alpha_i$ are the corresponding collective spin operators and $ a$  $(a^{\dagger}) $ is the annihilation (creation) operator for the electromagnetic mode. The first line in Eq.~\eqref{eq:h_edm} is the usual Dicke model,\mycite{BrandesPR2005} where $g$ denotes the coupling strength of a single dipole. This model describes well the collective interaction between $N$ dipoles and a common field mode in the weak coupling regime, $G=g\sqrt{N}\ll \omega_c$. 

At larger coupling strengths, the two additional spin-spin coupling terms in the second line of in Eq.~\eqref{eq:h_edm} must be taken into account. The first contribution $\sim S_x^2$ is the so-called depolarization or $P^2$-term. Its origin is related to the fact that $H_{\rm cQED}$ is derived in the dipole gauge,\mycite{Todorov2014,debernardis2018,debernardis2018b} where the canonical momentum variable of the electromagnetic mode is the displacement field, $\boldsymbol{D}= \varepsilon_0\boldsymbol{E} + \boldsymbol{P}$,\mycite{PhotonsAndAtoms} with $\boldsymbol{E}$ being the electric field and $\boldsymbol{P}\sim S_x$ the polarization density. Therefore, when expanding the electric field energy, $\sim \boldsymbol{E}^2\sim (\boldsymbol{D} - \boldsymbol{P})^2$, we obtain both the dipole-field interaction together with the accompanying $S^2_x$-term, which therefore should be interpreted as part of the field energy. In contrast, the last term in Eq.~\eqref{eq:h_edm} accounts for the actual dipole-dipole interactions, which exist independently of the dynamical mode. In free space we would simply obtain $J_{ij}\sim 1/|\boldsymbol{r}_i- \boldsymbol{r}_j|^3$, where $\boldsymbol{r}_i$ are the positions of the dipoles. However, in the presence of metallic boundaries, screening effects, etc., the actual dependence may be substantially modified,\mycite{debernardis2018} and can also be engineered to be infinite-ranged in circuit QED systems.\mycite{jaako2016,Bamba2016}

Equation~\eqref{eq:h_edm} shows that even at a minimal level, models of cavity QED involve collective interactions between spins and a bosonic mode as well as direct spin-spin interactions with different spatial dependencies. These terms are not completely independent of each other and in particular the strength of the $P^2$-term must match the dipole-field coupling to ensure consistency with basic electrodynamics. The static dipole-dipole interactions depend on the system and geometry under consideration and will in general introduce short-range interactions, which compete with the collective dipole-field coupling. Depending on the ratio of $g/\omega_c$ and the sign and strength of the couplings $J_{ij}$, different normal (paraelectric), superradiant (ferroelectric) and subradiant (anti-ferroelectric) phases can occur. This behaviour is illustrated in Figure \ref{Fig1:CavityQED}(b) for the simplified case $J_{ij}=J_0/N$, where due to symmetry the model can still be solved numerically for small and moderate numbers of dipoles.\mycite{debernardis2018} For short-range interactions, and depending on the geometry, different other types of phases may exists, but due to its computational complexity, little is still known about the ground and excited states of $H_{\rm cQED}$ in such general scenarios.

\section{Effective cavity QED models with trapped ions}

\begin{figure}[t]
    \begin{center}
        \includegraphics[width=\columnwidth]{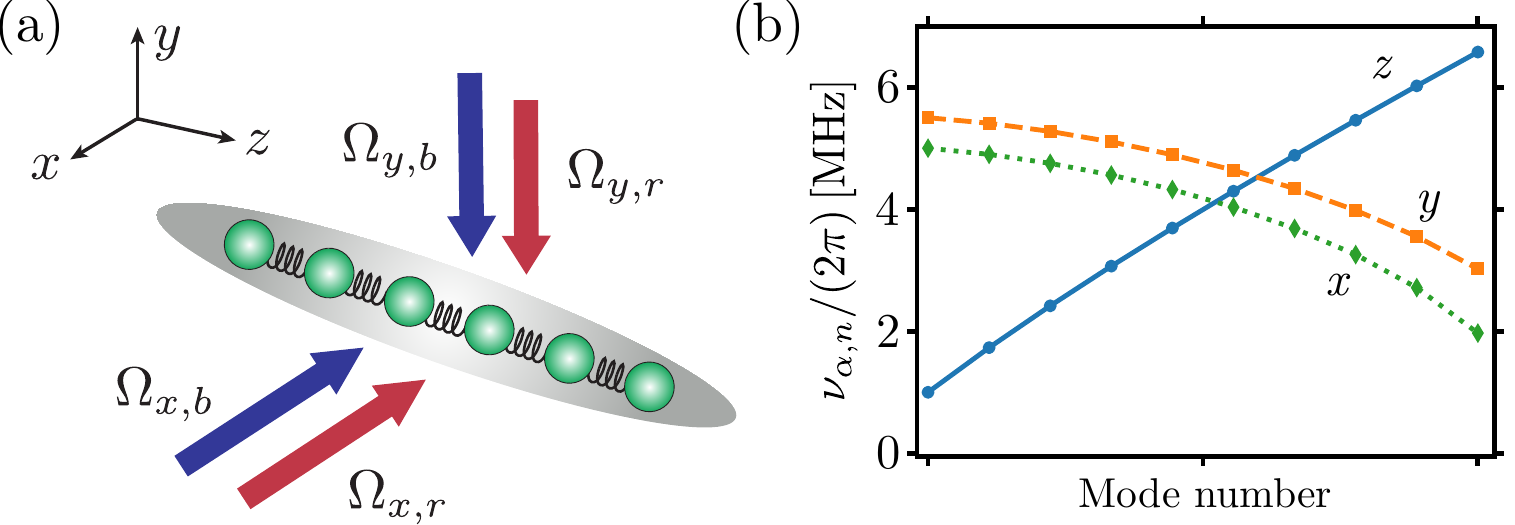}
        \caption{(a) A system of $ N $  trapped  ions in a linear Paul trap with the weakest confining potential along the $ z $ axis. The ions are driven with multiple blue- and red-detuned lasers along the other two directions to couple the internal states of the ions to different transverse phonon modes. (b) Example of a phonon spectrum of a chain of $N=10$ $^{40}$Ca$^+$ ions for an axial trap frequency of $ \nu_z/(2\pi)=1.0\,\mathrm{MHz} $ and transverse trapping frequencies of $ \nu_x/(2\pi) =  5.0\,\mathrm{MHz} $ and $ \nu_y/(2\pi) = 5.5\,\mathrm{MHz} $.}
        \label{Fig2:Setup}
    \end{center}
\end{figure}

For the implementation of $H_{\rm cQED}$ as an effective, but fully controllable, model we consider a system of $ N $ trapped ions in a linear Paul trap as shown in Figure \ref{Fig2:Setup}(a). At low enough temperatures the ions will arrange themselves in a one-dimensional (1D) chain and by writing the position of the $i$-th ion as $ \boldsymbol{r}_i=\boldsymbol{r}^{\,0}_i+\boldsymbol{u}_i$, we can linearize the motional dynamics around the equilibrium positions $\boldsymbol{r}^{\,0}_i$. The residual displacements $\boldsymbol{u}_i$ can be quantized and written in terms of a set of bosonic annihilation (creation) operators $b_{\alpha,n}$ ($b^\dag_{\alpha,n})$ as 
\begin{equation}
    \boldsymbol{u}_i=   \sum_{\alpha,n} \sqrt{\frac{1}{2M\nu_{\alpha,n}}} \xi_{n}^{\alpha}(i)\boldsymbol{e}_{\alpha} \left( b_{\alpha,n} +b_{\alpha,n}^\dag\right).
\end{equation}
Here $M$ is the mass of the ions and $\nu_{\alpha,n}$ and $\xi_{n}^{\alpha}$ denote the frequency and mode function of the $n$-th vibrational eigenmode in direction $\alpha$, respectively. A typical phonon spectrum is shown in Figure \ref{Fig2:Setup}(b) for the case $\nu_z< \nu_x < \nu_y$, where the $\nu_\alpha$ are the trapping frequencies along the three principal axes.

The ions are driven by two pairs of laser beams, which are slightly tuned to the red (r) and the blue (b) of  the transition frequency $\omega_{eg}$ between the long-lived electronic states $\ket{g}$ and $\ket{e}$. As indicated in Figure \ref{Fig2:Setup}(a), one pair of lasers is directed along the $x$-axis and the other pair along the $y$-axis and we assume that the beams are sufficiently broad such that they can be treated as plane waves.  The Hamiltonian of the whole ion chain is then given by
\begin{equation}\label{eq:Hion}
\begin{split}
    H &= \sum_{\alpha,n} \nu_{\alpha,n} b_{\alpha,n}^{\dagger}b_{\alpha,n}  + \sum_i \frac{\omega_{eg}}{2} \sigma_i^z \\
    &+  \sum_{\alpha,l,i} \dfrac{\Omega_{\alpha,l}}{2} \sigma^{x}_{i}  \left[    \ex^{\im \left[ \boldsymbol{k}_{\alpha,l}\cdot(\boldsymbol{r}^{\,0}_{i}+\boldsymbol{u}_i) - \omega_{\alpha,l}t \right]} + \textrm{H.c.} \right],
    \end{split}
\end{equation}
where $\omega_{\alpha,l}$ and  $\Omega_{\alpha,l} $ are the frequency and the  Rabi-frequency of laser $ l \in \{ r,\, b \} $ with wavevector $ \boldsymbol{k}_{\alpha,l}= k_{\alpha} \boldsymbol{e}_\alpha $. Since in the considered configuration the motion along the $z$-axis remains unaffected, we can restrict  $\alpha\in\{x,y\}$ and  assume $\ex^{\im \boldsymbol{k}_{\alpha,l}\cdot\boldsymbol{r}^{\,0}_{i}}\simeq 1$.

\subsection{Interaction engineering}
In the Lamb-Dicke regime, where the residual motion of the ions is small compared to the laser wavelength, we can expand the exponentials in Eq.~\eqref{eq:Hion} up to first order in the  displacements $\boldsymbol{u}_i$.\mycite{Leibfried2003} Under this approximation and by changing to a frame rotating with $\omega_{eg}$, the relevant laser-induced coupling between internal and external degrees of freedom reduces to 
\begin{align}\label{eq:h_ion_int}
    H_\Omega \simeq  \sum_{i,n,\alpha,l} \eta_{n}^{\alpha}\xi_{n}^{\alpha}(i) \dfrac{\Omega_{\alpha,l}}{2} \left[  \sigma^{+}_{i} (b_{\alpha,n} +  b^\dag_{\alpha,n})  \ex^{\im \delta_{\alpha,l}t } + \textrm{H.c.} \right],
\end{align}
where $\sigma_i^{\pm}=\sigma_i^x\pm i\sigma_i^y$, $\delta_{\alpha,l}=\omega_{\alpha,l}-\omega_{eg}$ and 
\begin{equation}
    \eta_{n}^{\alpha}=  k_\alpha \sqrt{\frac{1}{2M\nu_{\alpha,n}}}.
\end{equation}
As long as the couplings $\eta_{n}^{\alpha}\Omega_{\alpha,l}$ are small compared to the spacing between the vibrational modes, the detunings $\delta_{\alpha,l}$ can be chosen to resonantly enhance the interaction with a specific phonon mode, while the coupling to other modes as well as direct transitions between the internal states are strongly suppressed. This general scheme is frequently used in trapped ion systems to engineer different types of spin-phonon and spin-spin interactions and references to some of the relevant previous works in this field will be given in the following discussion. However, the cavity QED Hamiltonian \eqref{eq:h_edm} involves several different types of interactions with a finetuned relation between the coupling parameters. Therefore, here our goal is to show how these general techniques can be combined to engineer $H_{\rm cQED}$ with a large degree of control over all the parameters.

\subsubsection{Collective dipole-field coupling}
We first select  one of the vibrational modes to represent the photonic mode in the cavity QED model. Here we choose the transverse center-of-mass (COM) mode along the $x$-axis, which has a homogeneous mode profile and a frequency $\nu_{x,{\rm COM}}=\nu_x$. For the two laser beams along the $x$-direction we assume equal amplitudes, $\Omega_{x,l}=\Omega_x$, and detunings $\delta_{x,r/b} = \mp \nu_{x} + \Delta_{x,r/b}$, where $|\Delta_{x,r/b}|\ll \nu_{x} $. In the interaction picture with respect to the phonon modes and keeping only near-resonant terms we then obtain
\begin{align}
    H_\Omega \simeq &
    & \sum_{i} \dfrac{g}{2} \left[  \sigma^{+}_{i} \left(a \ex^{\im \Delta_{x,b} t } +  a^\dag \ex^{-\im \Delta_{x,r} t } \right) + \textrm{H.c.} \right],
\end{align}
where we identified $a\equiv b_{x,\rm COM}$ and $g\equiv \eta^x_{\rm COM}\xi_{\rm COM}^x\Omega_x$. We now define 
\begin{align*}
    &\omega_c = \dfrac{1}{2}(\Delta_{x,b} + \Delta_{x,r}),\\
    &\omega_0 = \dfrac{1}{2}(\Delta_{x,b} - \Delta_{x,r}),
\end{align*}
and after eliminating the time-dependence via the unitary transformation $U(t)=\ex^{-\im(\omega_c a^\dag a+\omega_0S_z)t}$ we obtain the effective Hamiltonian 
\begin{align}
    H_{\rm eff}^{(x)} \simeq  \omega_c a^\dag a  + \omega_0 S_z + g (a+a^\dag) S_x.
\end{align}
This and closely related schemes have already been discussed in many previous works for implementing effective Rabi- and Dicke models,\mycite{Dimer2007,Ballester2012,Zou2014,Pedernales2015,Puebla2016,Lv2017,SafaviNaini2018,Cohn2018} with the crucial benefit that the ratio between $g$, $\omega_c$ and $\omega_0$ is fully controlled by laser or microwave detunings, rather than by the bare physical parameters.

\subsubsection{The $P^2$-term}\label{sec:P2term}
The remaining terms in $H_{\rm cQED}$ contain direct spin-spin interactions $\sim \sigma_i^x\sigma_j^x$, with both constant and spatially varying prefactors. To implement  the collective $P^2$-term we use the two lasers along the $y$-direction to address the COM mode $b_{y,{\rm COM}}$ with $\nu_{y,{\rm COM}}=\nu_y$. By assuming equal Rabi frequencies $\Omega_{y,\ell}=\Omega_y$ and the detunings $\delta_{y,r/b} = \mp\mu + \omega_0$ we obtain the interaction-picture Hamiltonian
\begin{equation}\label{eq:h_spin_modes}
\begin{split}
    H_\Omega \simeq & \sum_{i,n} g_{n}^{y}\xi_{n}^{y}(i) \sin(\mu t) \\
    &\times\left(b_{y,n} \ex^{\im \nu_{y,n} t } + \mathrm{H.c.} \right) \left( \sigma_i^{+}\ex^{\im \omega_0 t} + \mathrm{H.c.} \right),
    \end{split}
\end{equation}
where $ g_{n}^{y} = \Omega_y\eta_{n}^y $. To realize pure spin-spin interactions, we consider the regime where $g_{n}^{y} \ll \mu - \nu_{y,n}$, such that the phonon modes are only virtually populated. The dynamics of the spins are then controlled by the spin Hamiltonian\mycite{Porras2004,friedenauer2008,Kim2009,Lin2011,islam2011,Dylewsky2016}
\begin{equation}\label{eq:h_spin_eff}
    H_{\rm eff}^{(y)} \simeq \sum_{i,j} \dfrac{D_{ij}}{4} \sigma_i^x\sigma_j^x.
\end{equation}
When the beat-note is chosen to be close to the COM motional frequency, $ \mu = \nu_y + \Delta_y $, with $|\Delta_y|$ being small compared to the mode spacing, the dipole couplings are approximately constant,
\begin{align}
   D_{ij} \simeq \frac{\left(g_{\rm COM}^{y} \xi_{\rm COM}^{y}\right)^2}{\Delta_y}.
\end{align}
This value can then be tuned to match $ g^2/\omega_c $ in order to reproduce the correct $ P^2$-term. Note that due to the presence of the local field $ \omega_0 $ in Eq.~\eqref{eq:h_spin_modes} terms proportional to $ b^{\dagger}_{y,n}b_{y,n}\sigma_i^{z} $ will be generated, which, however, are suppressed by a factor of $ \omega_0/(\mu - \nu_{y,n}) $ and can be neglected in the parameter regime of interest.\mycite{jurcevic2015}

\subsubsection{Dipole-dipole interactions}
To implement additional short-range dipole-dipole interactions, we generalize the scheme from above to laser beams with multiple modulation sidebands with slightly different frequencies. This can be accounted for by substituting in Eq.~\eqref{eq:h_ion_int}
\begin{equation}
    \Omega_{y} \ex^{\im\delta_{y,l} t} \rightarrow \sum_m \Omega_{y,m} \ex^{\im\delta_{y,l,m} t}.
\end{equation}
The resulting Hamiltonian is that of Eq.~\eqref{eq:h_spin_modes} with a sum over the different modulation sidebands. As long as the detunings between different modulation frequencies remain large, i.e., $\eta_{n}^{y} \Omega_m \ll |\delta_m - \delta_l|$, the resulting cross terms from lasers with different beat-note frequencies $ \delta_m $ are rapidly oscillating and can be neglected. The dynamics of the spins is then determined by an effective Hamiltonian as in Eq.~\eqref{eq:h_spin_eff}, where the generalized interaction matrix\mycite{Porras2004,friedenauer2008,Kim2009,Lin2011,islam2011,Dylewsky2016}
\begin{equation}\label{eq:Dij}
    D_{ij} = \sum_{m,n} 2\left({g_{n,m}^{y}}\right)^2\dfrac{\nu_{y,n}\xi_{n}^{y}(i)\xi_{n}^{y}(j)}{\mu_m^2 - \nu_{y,n}^2}
\end{equation}
can be engineered in a flexible manner by combining multiple near-resonant and/or far-detuned lasers. For example, by adding a laser which is far detuned from all modes, the coupling matrix will acquire an additional component which decays approximately as $\sim |i-j|^{-3}$ in the limit of very large detuning.\mycite{Porras2004,Kim2009} In contrast, when addressing one of the modes in the middle of the phonon band one obtains a coupling with an alternating sign, which typically leads to frustration and related phenomena.\mycite{Lin2011}

\subsection{Non-uniform couplings} 
The expression for $H_{\rm cQED}$ as given in Eq.~\eqref{eq:h_edm} is based on the usual assumption that the field profile is homogeneous over the extent of the ensemble of dipoles. However,  the model can easily be generalized to situations where the coupling strength $g_i$ is different for each dipole, by making also the corresponding substitution for the $P^2$ term, 
\begin{equation}
\frac{g^2}{\omega_c}S_x^2 \rightarrow   \sum_{i,j} \frac{g_i g_j}{4\omega_c} \sigma_i^x \sigma_j^x.
\end{equation}
The interaction engineering schemes discussed above allow the implementation of such non-uniform models by considering non-uniform mode profiles for the driving lasers, $\Omega_{x/y}\rightarrow \Omega_{x/y}({\bf r}^0_i)$. Since in our scheme we use the nearly uniform COM modes for engineering both the dipole-field coupling and the $P^2$-term, the correct relation between the two terms is automatically guaranteed when the same mode profiles for the lasers along the $x$ and the $y$ direction are assumed. Therefore, with the use of spatial light modulators or other experimental techniques, arbitrary mode profiles $g_i$ can be engineered while still retaining physically consistent models. However, for concreteness we will focus on the homogeneous case, $g_i=g$, in the remainder of the discussion.

\subsection{Accessible parameter regimes}
In summary, by addressing different phonon branches of the ion chain, both collective spin-photon and spin-spin interactions can be engineered independently. Therefore, by combining both schemes, $H_{\rm cQED}=H_{\rm eff}^{(x)}+H_{\rm eff}^{(y)}$, we obtain a class of cavity QED Hamiltonians with a dipole-dipole interaction matrix 
\begin{equation}\label{eq:Jij}
    J_{ij} = D_{ij} - \frac{g^2}{\omega_c},
\end{equation}
where the $D_{ij}$ are given by Eq.~\eqref{eq:Dij}. While in theory this approach provides full control over all relevant model parameters, the hierarchy of frequency scales and the single-mode addressability assumed in the derivation of the effective interactions still impose practical limitations on the accessible parameter regimes.

\subsubsection{Ultrastrong coupling regime}\label{ssec:USCParameters}
\begin{figure}[t]
    \begin{center}
        \includegraphics[width=\columnwidth]{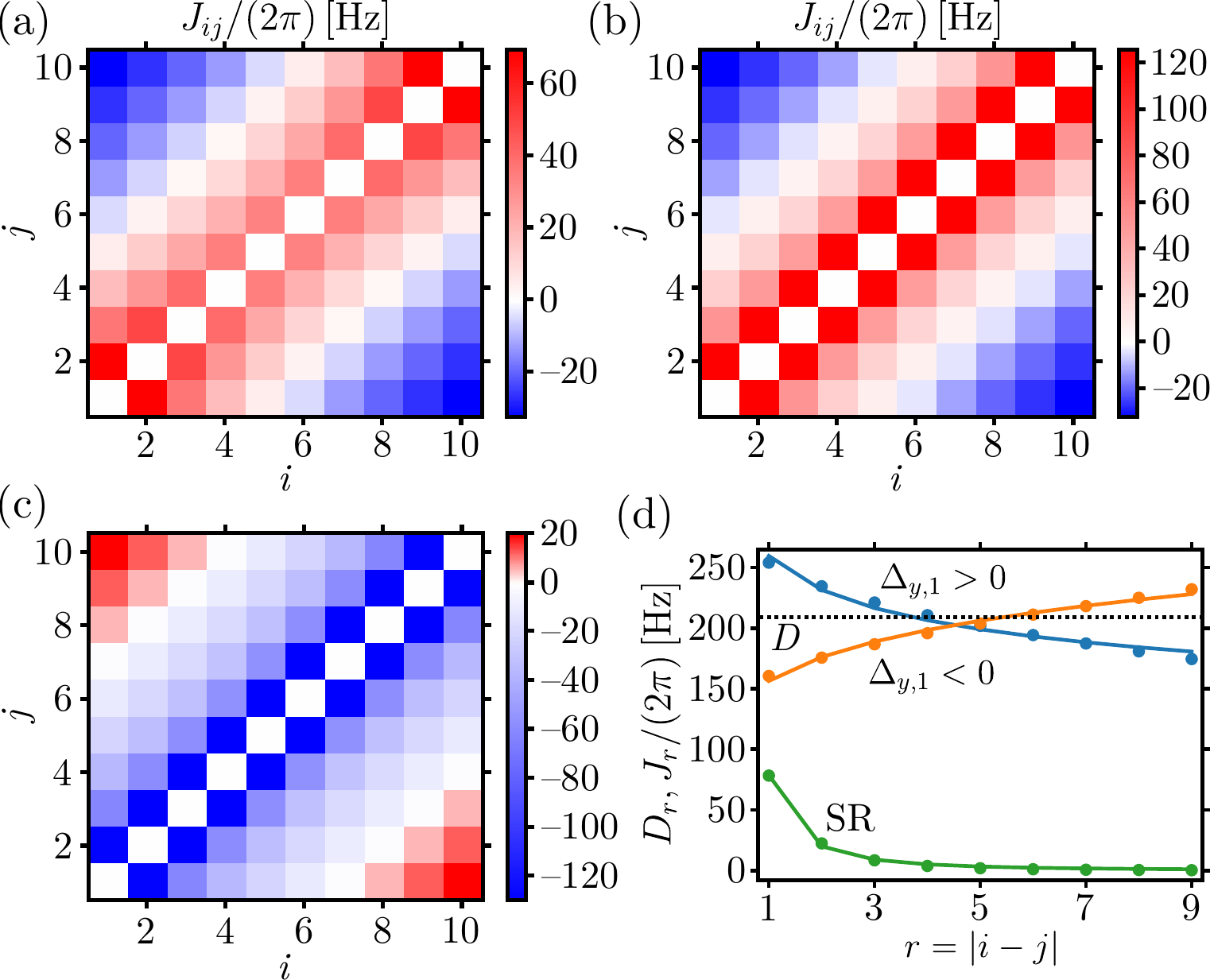}
        \caption{Dipole-dipole interaction matrices $J_{ij}$ for $N=10$ ions and the system parameters detailed in Sec.~\ref{ssec:USCParameters}. (a) The noninteracting case, where finite couplings $J_{ij}$ only arise from the residual variations of the matrix $D_{ij}$. (b) The case of repulsive dipole-dipole interactions. (c) The case of attractive dipole-dipole interactions. (d) Plot of the  distance dependence of the coupling matrix elements $\bar D_{|i-j|}$ for a single pair of lasers (upper two lines), where the bar denotes the average over all ions in the chain. The lower line shows the same distance dependence of $J_{r = |i - j|}$ when only the second, far-detuned pair of lasers is present. A fit $J_{r}\sim 1/r^\alpha$ yields a value of $\alpha\approx 2$. The black dotted line indicates the value of $ D $ given in the text.
        }
        \label{Fig3:dd_couplings}
    \end{center}
\end{figure}

As a specific example we consider a chain of $N=10$ trapped $^{40}$Ca$^+$ ions with a phonon spectrum as shown in Figure \ref{Fig2:Setup}(b). In this case the relevant Lamb-Dicke parameter is $\eta_{\rm COM}^{x} = 0.043$ and for $\Omega_x = 2\pi\times 15.4\,\mathrm{kHz}$, $\Delta_{x,b} = 2\pi\times 0.41\,\mathrm{kHz}$, and $\Delta_{x,r} = 0 $ we obtain $\omega_c = \omega_0 = 2\pi\times 0.21\,\mathrm{kHz} $ and a coupling parameter of $g/\omega_c=1$. For the implementation of the $ P^2$-term we follow the scheme in Sec.~\ref{sec:P2term} and use two lasers with Rabi-frequency $\Omega_{y,1}=2\pi\times 139\,\mathrm{kHz}$ to drive the COM mode with detuning $\Delta_{y, 1} = 2\pi\times 14\,\mathrm{kHz}$. For $\eta^y_{\rm COM} = 0.041$ this results in a collective $S_x^2$-coupling of strength $g^2/\omega_c = D = 2\pi\times 0.21\,\mathrm{kHz}$. Since in a real trap the mode function is not completely homogeneous and the laser will also weakly couple to all other $y$-modes, the exact evaluation of the coupling matrix $D_{ij}$ in Eq.~\eqref{eq:Dij} will result in small spatial variations, $D_{ij}\sim |i-j|^{-0.16}$. The resulting residual dipole-dipole interactions, $J_{ij}$, are plotted in Figure \ref{Fig3:dd_couplings}(a). On average, each dipole feels a residual field $\bar J_i=\sum_{j\neq i} J_{ij}$, with a variance $(\Delta \bar J)^2=\sum_i (\bar J_i)^2/N$ across the chain. For the current set of parameters $\Delta \bar J/\omega_0 \approx 0.19 $, and to a good approximation the dipoles can be considered non-interacting. Note that imperfections in $D_{ij}$ scale with $\sim g^2$ and become negligible for weaker couplings.

To add additional short-range dipole-dipole interactions in a controlled manner, we now consider a second pair of lasers along the $y$-axis with strength $\Omega_{y,2} = 2\pi \times 1.0\,\mathrm{MHz}$ and detuning $\Delta_{y,2} = 2\pi \times 1.7\,\mathrm{MHz}$. In this far-detuned limit the coupling to all the phonon modes is roughly the same and the resulting dipole-dipole couplings scale approximately as $J_{ij} \approx J_0/|i-j|^{\alpha}$, where $\alpha\simeq 1.98$ and $J_0 = 2\pi \times 80\,\mathrm{Hz}$, see Figure \ref{Fig3:dd_couplings}(d). In this case the total mean field $\bar J \approx 2\pi \times 0.20\,\mathrm{kHz}$ is comparable to $\omega_0$. The exact coupling matrix $J_{ij}$, including the residual imperfections from the $P^2$-term, is shown in Figure \ref{Fig3:dd_couplings}(b). For this simple driving scheme, the value of $\alpha\lesssim 2$ is limited by the ratio between the phonon bandwidth and the detuning. For a 1D chain the interactions can be considered as mid-range.\mycite{dutta2001} However, since the second laser must be detuned far to the blue, $\Delta_{y,2} >0$, the effective interactions are necessarily repulsive. To implement an equivalent model with attractive interactions, we can simply invert the sign of all the other terms in $H_{\rm cQED}$,\mycite{jurcevic2017} which can be done by replacing $\Omega_x\rightarrow -\Omega_x$ and changing the sign of the detunings, $\Delta_{x,r/b}$ and $\Delta_{y, 1}$. As a result we obtain the model $-H_{\rm cQED}$ with $J_0<0$. For the purpose of quantum simulation, the overall minus sign is unimportant. The plot in Figure \ref{Fig3:dd_couplings}(c) shows the resulting coupling matrix for $\Delta_{y,1} = -2\pi \times 11\,\mathrm{kHz}$ and $ \Omega_{y, 1} = 2\pi \times 112\,\mathrm{kHz} $, which, apart from the sign, leads essentially to the same effective parameters as above.

In summary, this example shows that trapped ions can be used to engineer few-body cavity QED models with coupling parameters $g/\omega_c\sim {\rm O}(1)$ and absolute frequency scales of a few hundreds of Hz. This is still fast compared to
simulation times of tens of milliseconds available in state-of-the-art trapped-ion experiments.\mycite{jurcevic2015,jurcevic2014,maier2019}

\subsubsection{Non-perturbative regime}
In the previous example the collective coupling $G=\sqrt{N}g$ already exceeds the cavity frequency by a factor of three. In the recent literature,\mycite{forndiaz19,kockum19} this regime is very generally called the DSC regime, without distinguishing between the collective and the single-dipole coupling constant. However, as indicated in the phase diagram in Figure \ref{Fig1:CavityQED}(b) and discussed in more detail in Ref.~\cite{debernardis2018}, significant non-perturbative changes in the physical properties of the cavity QED system are only expected beyond a value of $g/\omega_c\approx 2-3$ of the single-dipole coupling parameter, approximately independent of $N$. For simulating this regime, two main difficulties arise. First of all, by assuming a fixed value of $D=g^2/\omega_c\approx 2\pi\times 200\,\mathrm{Hz}$ as above, the frequencies $\omega_0\approx \omega_c \approx 2\pi \times 20\,\mathrm{Hz}$ must be reduced by a factor of about ten to reach these high values of the coupling parameter. Second, the reduced value of $\omega_0$ also means that any residual deviations of the actual coupling matrix, $\Delta \bar J/\omega_0 \sim  2 $, have now a much stronger impact on the bare model. These problems will in general become worse for larger $N$, where the conditions for single-mode resolution become more stringent and lead to a competition between the time-scales of the simulation and the quality of the model, i.e. the level of control over the interaction matrix $D_{ij}$. However, this is not a fundamental limitation and in particular for small and moderate numbers of ions, there are still many interesting effects that can be explored under those constraints.

\section{Examples}\label{sec:examples}
The parameters estimated above show that systems of trapped ions can be used to simulate otherwise unaccessible parameter regimes in cavity QED. In this section we discuss two basic examples, which also illustrate different measurement techniques that one can apply to extract interesting information about this system. 

\subsection{Few-dipole excitation spectrum}\label{ssec:spectrum}
As a first example we consider  the measurement of the excitation spectrum of a few-dipole cavity QED system in the parameter regime $g/\omega_c \lesssim 1$. The USC regime of cavity QED can be identified in the cavity excitation spectrum, as the region where the splitting between the two polariton modes, $\Delta \omega$, starts to deviate from the initial linear scaling $\Delta \omega \simeq G$. 
In typical  experiments in the optical and THz regime the condition $G\sim \omega_c$ is only accessible with a very large number of dipoles, where $g/\omega_c\ll1$ and only linearized collective excitations can be probed.\mycite{forndiaz19,kockum19} Corrections to this linear spectrum are expected to become observable for $N\lesssim 10$, \mycite{Todorov2014} but reaching this regime presents a notable experimental challenge. Superconducting circuits can more easily enter the USC regime, but the condition $g\simeq \omega_c$ has so far only been achieved with single flux qubits, due to the complexity of controlling and measuring multiple such devices. Finally, in all natural cavity QED systems, the dipole-dipole interactions are usually fixed or difficult to control. All these limitations are absent in our trapped-ion quantum simulator.

To measure the few-dipole excitation spectrum, the ions are initialized in state $\ket{g}$ and the photon mode is cooled to its ground state. Then, all the coupling terms are gradually increased from zero to their final value such that the system is adiabatically prepared in the ground state $\ket{G}$ of $H_{\rm cQED}$. Finally, a weak perturbation of the form $H_{p}(t)\sim A e^{i\omega t} + A^\dag e^{-i\omega t}$ is applied for a time $T_{p}$. In the limit $T_{p}\rightarrow \infty$, the amount of excitations created by such a perturbation will be proportional to the excitation spectrum 
\begin{equation}\label{eq:Spectrum}
    S(\omega)=  {\rm Re} \int_{0}^{\infty} \id \tau \langle A(\tau)A^{\dagger}(0) \rangle_{\rho_0} \ex^{\im \omega \tau},
\end{equation}
where the average is take over the actual state $\rho_0\approx \ket{G}\bra{G}$ after the adiabatic preparation. In practice, a  measurement of the expectation value $\langle A^\dag A\rangle$ before and after the applied perturbation will already provide an accurate estimate of $S(\omega)$ for finite $T_{p}$. For $A\equiv a$ this procedure provides a measurement of the cavity spectrum. In the following we focus instead on the case $A\equiv \sigma_3^-$, where $S(\omega)$ also contains information about the so-called dark polariton states, which are excitations of the dipoles that are decoupled from the cavity mode.

\begin{figure}[t]
    \begin{center}
        \includegraphics[width=\columnwidth]{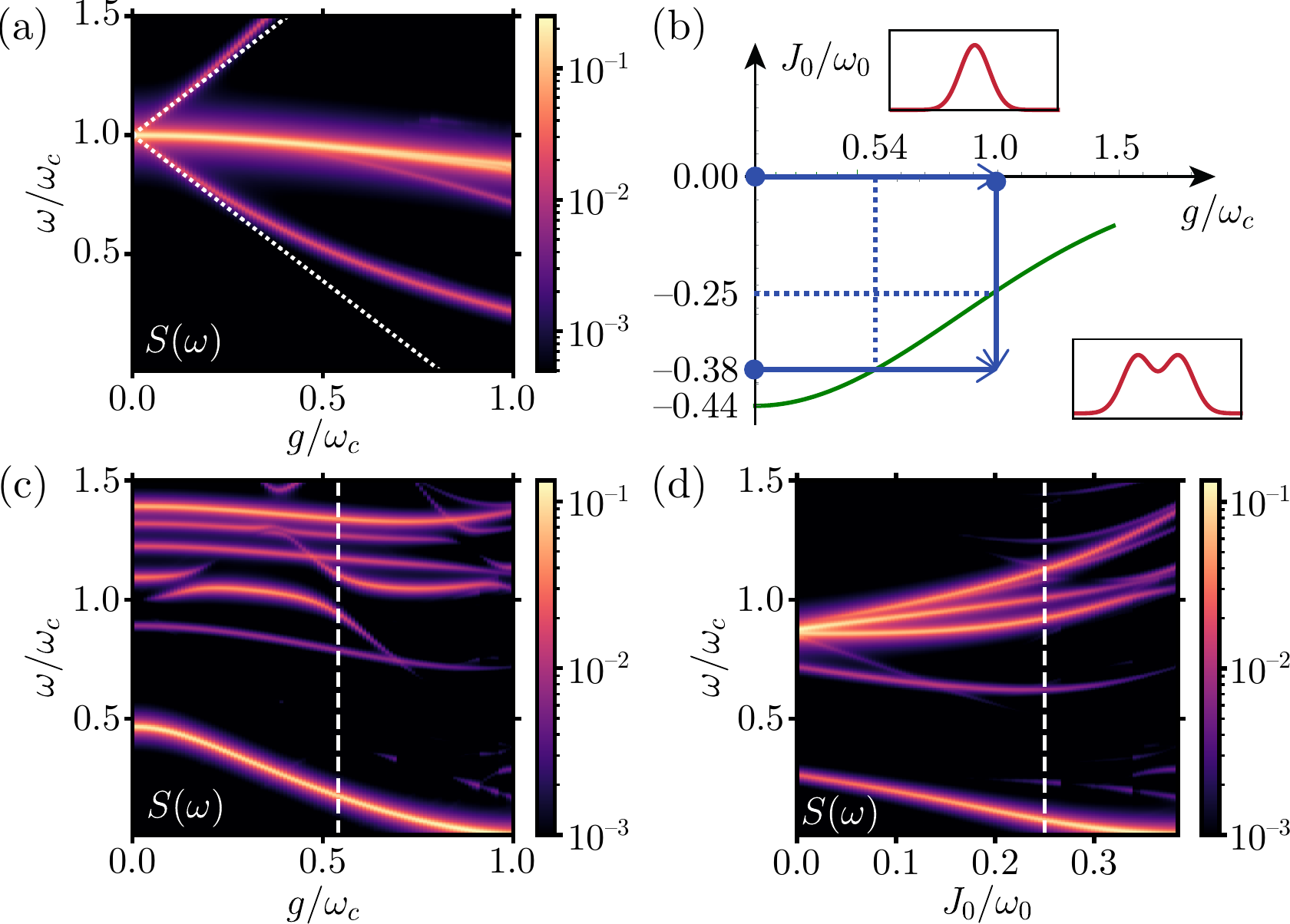}
        \caption{(a) Simulation of the excitation spectrum of a cavity QED system with $N=6$ non-interacting dipoles, $J_0=0$, and $\omega_0=\omega_c$. The dotted lines indicate the predictions from the Dicke model for the frequencies $\omega_{\pm}=\pm G/2$ of the two bright polariton modes. (b) Sketch of the phase boundary (green line) between the normal and the superradiant (ferroelectric) phase. As depicted by the two insets, across this boundary the distribution $p(m_x)$ changes from a single- to a bimodal distribution. The blue arrows indicate the range of the parameter sweeps in (a), (c) and (d). The plots in (c) and (d) show the variation of the excitation spectrum when the system is tuned across this phase boundary, in (c) for a fixed value of $ J_0/\omega_c \simeq -0.38 $ and in (d) for a fixed value  of $ g/\omega_c = 1 $. In both plots the dashed vertical line indicates the location of the phase boundary plotted in (b).}  
              \label{fig4:Spectrum}
    \end{center}
\end{figure}

Figure \ref{fig4:Spectrum} shows the numerically simulated result of such an experiment for the case of $N=6$ ions but otherwise similar parameters as discussed in Sec.~\ref{ssec:USCParameters}. Since from these simulations we found that a fully adiabatic preparation of the ground state requires a too long time of hundreds of milliseconds, 
we use a non-adiabatic bang-bang scheme, similar to what has been used previously.\mycite{Cohn2018,viola1998,balasubramanian2018} With this procedure detailed in App.~\ref{app:bangbang} the ground state can be prepared with a fidelity of $\mathcal{F}=\bra{G} \rho_0 \ket{G} \gtrsim 0.8$ in a time  $T_{\rm prep} \lesssim 7\,\mathrm{ms}$. Starting from this state, we use Eq.~\eqref{eq:Spectrum} to evaluate the excitation spectrum for different frequencies $\omega$, where we simply assume a common phenomenological decay rate of $\Gamma = 2\pi \times 4\,\mathrm{Hz}$ for all excited states, corresponding to an experimental runtime of $T_p\approx 40\,\mathrm{ms} $.

In Figure \ref{fig4:Spectrum}(a) the resulting spectrum is first plotted for non-interacting dipoles, where up to the residual imperfections described in Figure \ref{Fig3:dd_couplings}, $J_{ij} \approx 0$. For small $G$ one observes the expected Rabi-splitting $\Delta \omega\simeq G$ between the two bright polariton states 
\begin{equation}
    |\Psi_p^{\pm}\rangle = \frac{1}{\sqrt{2}}\left(  a^\dag \pm \frac{1}{\sqrt{N}} \sum_{i=1}^N \sigma_+^i \right) |G\rangle,
\end{equation} 
while other excitations of the dipoles are decoupled from the cavity and remain almost unaffected. Note that since we consider the response of a single dipole, the signal is sensitive to all excitation modes, but the overlap with the collective polariton modes is reduced by a factor $1/N$. Thus, the ability to see both collective and single-particle effects is a specifically interesting feature of the considered few-dipole regime. At large couplings the influence of the $P^2$-term is no longer negligible and the spectrum starts to deviate from the predictions of the usual Dicke model.\mycite{Ciuti2005,Todorov2012,Todorov2010,Maissen2014,Zhang2016} In particular, the frequency of the lower polariton mode stabilizes at a non-zero value for all couplings.\mycite{Ciuti2005,Todorov2012} A somewhat unexpected observation is the downward shift of the dark polariton modes, which is not predicted by a purely linear theory. It arises from the fact that the ground state energy $ E_G(g) $ 
increases with increasing $g$. When one of the dipoles is now promoted to a decoupled mode, less energy is needed. Finally, a finite splitting between the dark modes indicates residual dipole-dipole interactions $J_{ij}$ due to a nonuniform matrix $D_{ij}$.

As a next step we switch to a system with strong attractive dipole-dipole interactions, $J_{ij}\approx J_0/|i-j|^\alpha$ and $J_0<0$. In this case the dipoles can undergo a transition into a ferroelectric state at a critical coupling $J_0^c$. For an infinite system and $\alpha=2$ a value of $J_0^c/\omega_0 \simeq -0.4$ is predicted.\mycite{jaschke2017} In the presence of the cavity mode this value is expected to decrease as
\begin{equation}
    J_0^c(g)\approx J_0^c(g=0) \ex^{-g^2/(2\omega_c^2)},
\end{equation}
due to the dressing of the dipoles by virtual photons.\mycite{debernardis2018} Of course, for the considered small number of dipoles, $N=6$, there is only a smooth crossover between the normal and the ferroelectric phase. However, the two phases can still be distinguished by looking at the probability distribution $p(m_x) = \bra{G}\mathbbm{P}_{m_x}\ket{G}$, where $\mathbbm{P}_{m_x} = \sum_s \mathbb{P}_{s,m_x}$ and $ \mathbb{P}_{s,m_x} $ is the projector on states with $S_x\ket{\psi} = m_x\ket{\psi}$ and total spin $ s $. The phase boundary can then be defined as the line, where this function changes from a single to a bi-modal distribution. In Figure \ref{fig4:Spectrum}(b) this boundary is sketched for the current model parameters and for different values of the light-matter coupling $g$.

We see that within the accessible parameter range the transition line can be crossed in two different ways: Either in the conventional sense, by increasing $|J_0|$, or by keeping $|J_0|<|J_0^c|$ fixed, but varying the coupling to the cavity. The corresponding spectra are shown in Figure \ref{fig4:Spectrum}(d) and (c). We see that in both cases the frequency of the lowest excited mode goes to zero around the expected transition point, where one must keep in mind that for a finite system the energy gap at $J_0^c(g)$ remains finite. The difference between the two tuning schemes is clearly visible in the excited states. In the first case, avoided crossings around $\omega_c$ indicate strong hybridization with the cavity mode, while the same features are absent when only $J_0$ is varied.

We emphasize that the strong reduction of the lower polariton frequency observed in Figure \ref{fig4:Spectrum}(c), which indicates the transition from the normal to the ferroelectric phase, is a true non-perturbative effect. It arises from to the renormalization of the dipole frequency $\omega_0$, while  the strength of static dipole-dipole interactions remains fixed. This is in contrast to the celebrated superradiant phase transition in the Dicke model.\mycite{BrandesPR2005} From Eq.~\eqref{eq:h_edm} we see that when interpreted in the context of cavity QED, the Dicke model corresponds to the case of a ferroelectric ensemble of dipoles with $J_{ij} = -g^2/\omega_c$. As a consequence, when increasing $g$ in this model, the system undergoes a regular ferroelectric phase transition, where the coupling to the cavity mode only introduces minor modifications.\mycite{debernardis2018}

\subsection{Subradiant ground state}
While already in the regime $g/\omega_c\lesssim 1$ first non-perturbative corrections are observable, the ground state is still determined primarily by the competition between $\omega_0$ and $J_{ij}$ and the influence of the cavity mode is mainly seen in the excited states. This changes drastically in the regime $g/\omega_c > 2$, where apart from the renormalization of the transition frequency, the cavity also induces effective anti-ferroelectric interactions, $H_{AF}\simeq J_{AF} (S_x^2-\boldsymbol{S}^2)$, where $J_{AF}=\omega_c\omega_0^2/(2g^2) > 0$.\mycite{jaako2016} For small $|J_0|$, these effective interactions compete with the short-range couplings and favor so-called subradiant ground states with completely anti-aligned dipoles that are decoupled from the cavity mode.

\begin{figure}[t]
    \begin{center}
        \includegraphics[width=\columnwidth]{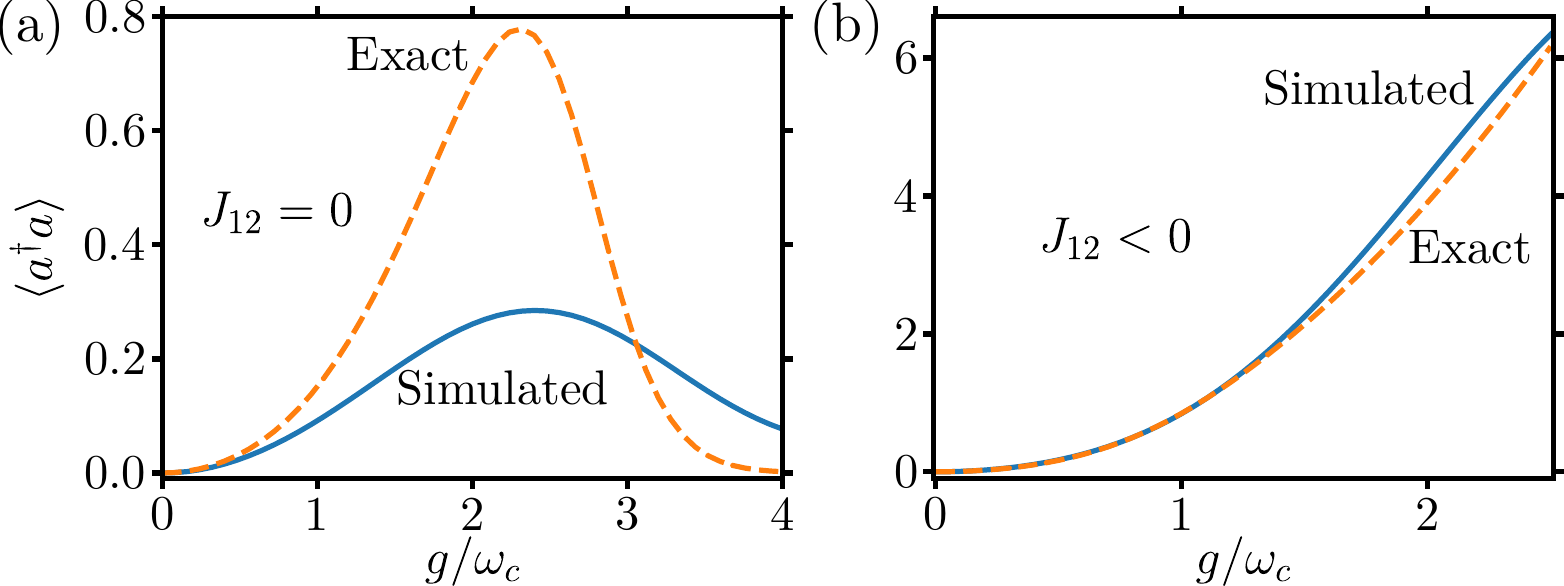}
        \caption{Adiabatic ground-state preparation in the DSC regime for the case of (a) $ J_{12} = 0 $ and (b) $ J_{12}/\omega_c = -3.5$ and $\omega_0=\omega_c$. The ground state is prepared by adiabatically turning on the dipole-field coupling $g$ and dipole-dipole interactions $J_0$ on a timescale of $T_{\rm prep} \lesssim 10\,\mathrm{ms} $, starting from the ground state of the noninteracting system.}
        \label{Fig5:GroundState}
    \end{center}
\end{figure}

The simplest experimental setting where this effect can be explored is the case of $N=2$ ions. In this case the matrices $D_{ij}=D$ and $J_{ij}=J_0$ have only one relevant entry, which allows us to relax some of the detuning constraints and consider values of $D = 2\pi \times 2\,\mathrm{kHz}$ and corresponding values of $\omega_{c,0} \approx 2\pi \times 100-500\,\mathrm{Hz}$ to access coupling parameters up to $g/\omega_c = 4$. For these parameters we plot in Figure \ref{Fig5:GroundState}(a) and (b) the expectation value of the photon number $\langle a^\dag a\rangle$ in the simulated ground state, $\rho_0$, for $J_0=0$ and $J_0/\omega_c=-3.5$. For small $g$ we see in both plots the expected increase of the photon number due to a hybridization between the dipoles and the photons. For the ferroelectric case, the photon number then increases rapidly after $g/\omega_c\approx 1$, which is the characteristic signature of a superradiant phase. In contrast, for non-interacting dipoles this trend turns around after $g\gtrsim 2\omega_c$ and the cavity returns back to its ground state for very large couplings. While for the considered preparation time $T_{\rm prep}\approx 10\,\mathrm{ms}$ the simulated photon number still differs from that of the true ground state, the characteristic maximum, which is the key signature for entering a subradiant ground state,\mycite{jaako2016,debernardis2018} is clearly visible. In a trapped ion quantum simulator this effect can be verified independently by performing a full tomography of the internal state. For the maximal coupling we find an overlap of ${\rm Tr}\{ \rho_0 |T\rangle\langle T|\}\approx 0.99$, where $|T\rangle =(|ee\rangle - |gg\rangle)/\sqrt{2}$ is the maximally entangled state that minimizes $H_{\rm AF}$.

\subsection{Validity of the effective model and decoherence}
All  the results presented in this sections are based on numerical simulations of the effective model, taking all imperfections of the coupling matrix $J_{ij}$ into account, but neglecting the weak admixture of other phonon modes. To ensure validity of the effective model, we have chosen parameters such that 
\begin{equation}
    \frac{(\eta_{\rm COM}^y)^2\Omega_y^2}{\Delta_y^2} N < 0.1.
\end{equation}
This means that the occupation of the COM mode, which is used to implement the $P^2$-term should be at most a few percent, $\langle b^\dag b\rangle < 0.1$. We have explicitly verified this estimate
by performing numerical simulations where the dynamics of the COM mode is included. In these simulations we do not see any significant changes in the dynamics of the system when compared to the effective one-mode model. 

In real experiments the system will also be affected by decoherence of the internal states and heating of the phonon modes. In all the examples discussed above the time for preparing the ground state, $T_{\rm prep}$, is chosen to be at most $ 10\,\mathrm{ms} $. In state-of-the-art ion traps the heating rates can be as low as 1-10 quanta per second,\mycite{brownnutt2015,schindler2013} meaning that the number of added phonons during the simulations is less than ten percent. In addition, for $g/\omega_c>1$ the spin states start to decouple from the oscillator mode and, thus, the internal state is even less affected by heating. A remaining source of error is  the dephasing of the internal states by magnetic field fluctuations or laser phase noise. For the adiabatic ground state preparation scheme, the ions are initially in an eigenstate of $\sigma_z$ and, therefore, dephasing is only relevant in the final part of the protocol. A master equation simulation, including also a heating rate of 10 quanta per second for the phonon mode, of the ground state preparation protocol of Figure \ref{Fig5:GroundState} for $ g/\omega_c = 4 $ shows that for a dephasing time as low as $ T_2=10\,\mathrm{ms} $, the fidelity of the final state only changes by $\Delta \mathcal{F}= 0.15$ and by only $\Delta \mathcal{F}= 0.02$ for realistic dephasing times of $T_2 = 100\,\mathrm{ms}$.\mycite{monz2011} These findings are consistent with other quantum simulation experiments, where simulation times of $>50$ ms have been demonstrated.\mycite{jurcevic2014, maier2019}

\section{Discussion and Conclusions}\label{sec:conc}
In summary, we have presented a comprehensive analysis on the suitability of ion traps for simulating the extended Dicke model~\eqref{eq:h_edm}, which captures the essential non-perturbative effects of cavity QED systems in the USC and DSC regime. Compared to real cavity or circuit QED systems, such simulators provide a flexible way to tune independently the coupling between the dipoles to a dynamical cavity mode and direct electrostatic interactions. The two examples discussed in more detail in Sec.~\ref{sec:examples} illustrate different possibilities for exploring characteristic signatures of non-perturbative light-matter interactions in the ground and excited states of few-body cavity QED systems. 

The analysis of this work has been restricted to a small number of ions, where adiabatic and non-adiabatic ground-state preparation schemes can still be benchmarked by a comparison with exact numerics. Similar to many closely related proposals,\mycite{Dimer2007,Ballester2012,Zou2014,Pedernales2015,Puebla2016,Schneeweiss2018,aedo2018} in this case the simulation aspect primarily lies in the ability to study coupling regimes that are fundamentally not accessible in systems of atomic or molecular dipoles. However, in principle, the same ideas can be generalized to several tens of ions or multiple cavity modes in order to explore non-perturbative effects far beyond the reach of classical simulation capabilities. From our numerical studies we find that the main practical difficulty in doing so arises from the collective $ P^2 $-term, $\sim g^2/\omega_c S_x^2$, which becomes the dominant contribution in the regime $g/\omega_c\gtrsim 1$. This term is implemented by selectively addressing the center-of-mass mode, which becomes more and more difficult as the number of ions increases and significantly prolongs the experimental timescales. This feature makes the extended Dicke model a particular challenge for trapped-ion systems and other quantum-simulation platforms.  

However, we envision that with improved motional heating and spin coherence times in future ion traps, simulation timescales of several seconds will become possible. It also has been shown\mycite{GarciaRipoll2005,teoh2019} that with full single-site addressability, stroboscopic techniques and numerical optimization the design of the coupling matrix $D_{ij}$ can be considerably improved to reduce residual imperfections while retaining a high coupling strength. Therefore, with further experimental and theoretical work along these lines, also the simulation of non-perturbative effects in cavity QED systems with tens of dipoles and multiple modes is achievable, where currently neither analytic predictions nor numerical simulations are available.  



\acknowledgments
We thank Ana Maria Rey for valuable discussions. This work was supported by the Austrian Science Fund (FWF) through Grant No. P 31701-N27 and DK CoQuS, Grant No. W 1210, and by an ESQ Discovery Grant of the Austrian Academy of Sciences (\" OAW). J. J. Garc\' ia-Ripoll acknowledges support from AEI Project PGC2018-094792-B-I00, CSIC Research Platform PTI-001, and CAM/FEDER Project No. S2018/TCS-4342 (QUITEMAD-CM).

\appendix
\section{Bang-bang state preparation}\label{app:bangbang}
In this appendix we give a short overview of the bang-bang state preparation scheme\mycite{Cohn2018,balasubramanian2018,viola1998} used in the numerical simulations presented in  Sec.~\ref{ssec:spectrum}. Let us consider a Hamiltonian that depends on a tunable parameter $ \lambda $, $ H = H(\lambda) $. The goal is to prepare the system in the ground state $ \ket{\psi_f} $ of $ H(\lambda) $ for some target value $ \lambda = \lambda_f $. This can be approximately achieved as follows. In the beginning of the protocol, the parameter $ \lambda $ has a value $ \lambda = \lambda_0 $, which is chosen in such a way that the system can easily be prepared in the ground state of $ H(\lambda_0) $. In a second step, the Hamiltonian is quenched to $ H(\lambda_1) $ with $ \lambda_1 \neq \lambda_0 $, and the system evolves for time $ T $ under the action of $ H(\lambda_1) $. Finally, at time $ T $ the tuning parameter is quenched to the target value $ \lambda_f $. One can then numerically optimize the intermediate value of the tuning parameters $ \lambda_1 $ and the time $T$ to obtain the largest overlap with the time evolved state $ \ket{\psi(T)} $ and the target state $ \ket{\psi_f} $.

We are interested in preparing ground states of the extended Dicke model for different values of $ g $ and $ J_0 $. The qubit frequency $ \omega_0 $ is chosen as the tunable parameter $ \lambda $ in this case. As an initial state we use the $ g = J_0 = 0 $ ground state $ \ket{n = 0}\otimes\ket{m_z = -N/2} $ with $ N $ the number of ions. Then for fixed $ g $ and $ J_0 $ the bang-bang protocol is performed. For each $ g $ and $ J_0 $, in general, a different intermediate $ \omega_0 $ and waiting time $ T $ are used to obtain the best overlap with the targeted ground state.

\end{document}